# Light tunneling inhibition and anisotropic diffraction engineering in two-dimensional waveguide arrays

Y. V. Kartashov,[1] A. Szameit,[2] V. A. Vysloukh,[3] and L. Torner[1]

[1]ICFO-Institut de Ciencies Fotoniques, and Universitat Politecnica de Catalunya, Mediterranean Technology Park, 08860 Castelldefels (Barcelona), Spain

[2]Physics Department and Solid State Institute, Technion, 32000 Haifa, Israel

[3]Departamento de Fisica y Matematicas, Universidad de las Americas – Puebla, 72820, Puebla, Mexico

We address two-dimensional waveguide arrays where light tunneling into neighboring waveguides may be effectively suppressed by out-of-phase harmonic modulation of the refractive index in neighboring waveguides at suitable frequencies. Genuine two-dimensional features, such as anisotropic diffraction engineering, diffraction-free propagation along selected directions in the transverse plane and tunneling inhibition for multi-channel vortices, are shown to occur.

    A periodic transverse modulation of the refractive index in optical materials is a powerful tool to control the propagation dynamics of light [1,2], allowing, e.g., engineering diffraction. Enhanced possibilities are made possible by bi-periodic modulations along both, transversal and longitudinal directions, affording a variety of new phenomena, such as diffraction-managed solitons [3,4], dragging of laser beams [5,6], periodic shape transformations or Rabi oscillations [7,8], parametric amplification of soliton swinging [9], just to name a few. Diffraction inhibition is also possible, as it was shown in periodically curved arrays [10-15], in arrays with oscillating widths of channels [16], or in lattices with longitudinally oscillating refractive index [14,17,18,19]. In this Letter we address light tunneling inhibition in two-dimensional (2D) honeycomb waveguide arrays [20]. In this specific configuration it is possible to realize an out-of-phase modulation of the refractive index in neighboring waveguides for the entire array, which results in the suppression of light tunneling. One can

engineer the diffraction by properly selecting clusters of out-of-phase or in-phase modulated waveguides and thus achieve, e.g. 1D diffraction in a 2D array. In addition, we show that tunneling inhibition is also possible for complex light patterns such as optical vortices.

We describe the propagation of cw radiation along the $\xi$-axis of a modulated waveguide array with the nonlinear Schrödinger equation for the dimensionless field amplitude $q$:

$$i\frac{\partial q}{\partial \xi} = -\frac{1}{2}\left(\frac{\partial^2 q}{\partial \eta^2} + \frac{\partial^2 q}{\partial \zeta^2}\right) - |q|^2 q - pR(\eta,\zeta,\xi)q. \qquad (1)$$

Here $\eta,\zeta$ and $\xi$ are normalized transverse and longitudinal coordinates, whereas $p = 12$ describes the refractive index modulation depth. The structure of the array is described by the function $R(\eta,\zeta,\xi)$, which is composed of Gaussian waveguides $\exp[-(\eta^2 + \zeta^2)/d^2]$ with centers $(\eta_k,\zeta_k)$ placed in the nodes of a honeycomb grid [Fig. 1(a)]. The waveguide width and the spacing between them are set to $d = 0.5$ and $s = 2$, respectively. The refractive index is modulated out-of-phase in neighboring waveguides, i.e. if in the central guide the refractive index oscillates as $1 + \mu\sin(\Omega\xi)$, where $\mu \ll 1$, in all adjacent waveguides it changes as $1 - \mu\sin(\Omega\xi)$. The value $\Omega$ represents the modulation frequency. Importantly, in the honeycomb array one can realize a configuration when each waveguide is completely surrounded by neighbors with out-of-phase longitudinal refractive index modulation. As the initial condition in all simulations of Eq. (1) we use $q|_{\xi=0} = Aw(\eta,\zeta)$, where $A$ is the amplitude and the function $w(\eta,\zeta)$ describes the profile of the linear guided mode of an isolated waveguide with $\max[w(\eta,\zeta)] = 1$. It is worth stressing that the longitudinally modulated lattices considered here are not equivalent to curved ones [13,14]. While the latter exhibit a single-waveguide unit cell, out-of-phase modulated waveguide arrays consists of binary unit cells, therefore splitting the first propagation constant band into two sub-bands. Hence, self-collimation in the second case requires a setting with two sub-lattices, which is provided by a honeycomb configuration.

Assuming only two coupled waveguides, the light switches periodically between them with a constant period $T_b$ provided that $A \to 0$ and $\mu = 0$. In our case we find $T_b = 67.17$. The same tunneling mechanism occurs in waveguide arrays ($\mu = 0$) yielding considerable spreading of light patterns with distance (see Fig. 3, upper row). This picture changes in modulated arrays ($\mu \neq 0$), where tunneling can be inhibited almost completely for any distance $\xi$ under appropriate resonant conditions (Fig. 3, lower row). Within the frame of the tight-binding approximation, adding a longitudinal modulation is equivalent to a reduction of the coupling constant $C \sim 1/T_b$ by the factor of $J_0(2\mu/\Omega)$ [19]. Coupling thus vanishes completely for $2\mu/\Omega = \nu_j$, with $\nu_j \approx 2.4, 5.5, ...$ being roots of the zero-order Bessel function. To demonstrate this feature, we show in Fig. 1(b) the distance-averaged power fraction trapped in the excited channel:

$$U_m = L^{-1} \int_0^L d\xi \int \int_{-s/2}^{s/2} |q(\eta,\zeta,\xi)|^2 \, d\eta d\zeta \Big/ \int \int_{-s/2}^{s/2} |q(\eta,\zeta,0)|^2 \, d\eta d\zeta \qquad (2)$$

as a function of the modulation frequency normalized by $\Omega_b = 2\pi/T_b$ at $L = 4T_b$ and $A = 0.01$. One can see that $U_m(\Omega)$ features several maxima corresponding to the respective resonances. The main resonance (at largest frequency $\Omega_r$) is always pronounced most, so that tunneling is strongly inhibited (Fig 3, second row). The frequency of the main resonance grows linearly with $\mu$ almost everywhere, except for the small vicinity of the point $\mu = 0$ where this dependence is parabolic and $\Omega_r/\Omega_b \to 0.5$ [Fig. 1(c)]. Interestingly the efficiency of tunneling inhibition [that is characterized by the resonant value $U_m(\Omega = \Omega_r)$] rapidly grows with $\mu$ and then saturates at $U_m \approx 0.92$ for $\mu > 0.08$. This efficiency can be further enhanced in waveguide arrays with larger $p$, where the confinement of the guided waveguide modes is stronger. For small longitudinal modulation depths ($\mu \sim 0.02$) inhibition becomes less efficient since $U_m$ remains small even in the primary resonance.

Importantly, a weak nonlinearity enhances tunneling inhibition. Figure 2(a) shows $U_m(\Omega)$ for small and moderate input amplitudes. The amplitude growth initially results in a considerable broadening of the resonance curve and a slight increase of $U_m(\Omega_r)$, whereas the resonance frequency does not change notably. The nonlinearity-induced broadening of the resonance curve may be so large that the width $\delta\Omega$ of the primary resonance [defined at the level of 70% from the maximal value $U_m(\Omega_r)$] becomes comparable with the resonance frequency [Fig. 2(b)]. Notice that for moderate $A^2$ the width of the resonance grows almost linearly with the input power. A further growth of the input amplitude first results in nonlinearity-induced delocalization (analogous to that observed in one-dimensional systems [4,19]), and then to re-localization at even higher amplitudes as expected from a soliton-like viewpoint.

It should be pointed out that in two-dimensional modulated waveguide arrays nontrivial diffraction control is possible. It might be realized by dividing the entire array into clusters where in each cluster the refractive index of adjacent guides oscillates in-phase, but in waveguides belonging to different clusters it oscillates out-of-phase. In honeycomb arrays featuring three principal axes one can have waveguides oscillate in-phase in the direction parallel to the principal axis, but out-of-phase in the direction perpendicular to it. In this case light beams will diffract along the selected principal axis, while in the perpendicular direction the diffraction will be inhibited at the modulation frequency $\Omega = \Omega_{\rm r}$. This results in essentially 1D anisotropic diffraction in an intrinsically 2D array (Fig. 4).

Finally, it should be stressed that tunneling inhibition in honeycomb arrays is possible not only for simplest excitation of a single-channel, but also for multiple-channel excitations. This enables simultaneous diffractionless transmission of several beams launched in different locations in the array. As a particular example we illustrate in Fig. 5 the evolution of a linear vortex beam residing on six channels of the structure and carrying unit topological charge. In an unmodulated array such a beam undergoes strong diffraction [Figs. 5(a)-5(c)], while in a modulated array under resonance conditions the vortex preserves its characteristic six-lobed intensity distribution and, more importantly, its topological charge at any distance [Figs. 5(d)-5(f)]. Such vortices survive not only in linear, but also in nonlinear regimes [Figs. 5(g)-5(i)].

Summarizing, we showed that light tunneling might be inhibited in 2D honeycomb waveguide arrays with out-of-phase modulation of the refractive index in neighboring channels. Such structures allow tailoring of the diffraction properties and diffractionless transmission of complex multiple-channel light patterns.

# References with titles

1. F. Lederer, G. I. Stegeman, D. N. Christodoulides, G. Assanto, M. Segev, and Y. Silberberg, "Discrete solitons in optics," Phys. Rep. **463**, 1 (2008).

2. Y. V. Kartashov, V. A. Vysloukh, and L. Torner, "Soliton shape and mobility control in optical lattices," Prog. Opt. **52**, 63 (2009).

3. M. J. Ablowitz and Z. H. Musslimani, "Discrete diffraction managed spatial solitons," Phys. Rev. Lett. **87**, 254102 (2001).

4. A. Szameit, I. L. Garanovich, M. Heinrich, A. Minovich, F. Dreisow, A. A. Sukhorukov, T. Pertsch, D. N. Neshev, S. Nolte, W. Krolikowski, A. Tünnermann, A. Mitchell, and Y. S. Kivshar, "Observation of diffraction-managed discrete solitons in curved waveguide arrays," Phys. Rev. A **78**, 031801 (2008).

5. Y. V. Kartashov, L. Torner, and D. N. Christodoulides, "Soliton dragging by dynamic optical lattices," Opt. Lett. **30**, 1378 (2005).

6. C. R. Rosberg, I. L. Garanovich, A. A. Sukhorukov, D. N. Neshev, W. Krolikowski, and Y. S. Kivshar, "Demonstration of all-optical beam steering in modulated photonic lattices," Opt. Lett. **31**, 1498 (2006).

7. Y. V. Kartashov, V. A. Vysloukh, and L. Torner, "Resonant mode oscillations in modulated waveguiding structures," Phys. Rev. Lett. **99**, 233903 (2007).

8. K. Shandarova, C. E. Rüter, D. Kip, K. G. Makris, D. N. Christodoulides, O. Peleg, and M. Segev, "Experimental observation of Rabi oscillations in photonic lattices," Phys. Rev. Lett. **102**, 123905 (2009).

9. Y. V. Kartashov, L. Torner, and V. A. Vysloukh, "Parametric amplification of soliton steering in optical lattices," Opt. Lett. **29**, 1102 (2004).

10. S. Longhi, M. Marangoni, M. Lobino, R. Ramponi, P. Laporta, E. Cianci, and V. Foglietti, "Observation of dynamic localization in periodically curved waveguide arrays," Phys. Rev. Lett. **96**, 243901 (2006).

11. R. Iyer, J. S. Aitchison, J. Wan, M. M. Dignam, C. M. de Sterke, "Exact dynamic localization in curved AlGaAs optical waveguide arrays," Opt. Express **15**, 3212 (2007).

12. F. Dreisow, M. Heinrich, A. Szameit, S. Doering, S. Nolte, A. Tünnermann, S. Fahr, and F. Lederer, "Spectral resolved dynamic localization in curved fs laser written waveguide arrays," Opt. Express **16**, 3474 (2008).
13. I. L. Garanovich, A. Szameit, A. A. Sukhorukov, T. Pertsch, W. Krolikowski, S. Nolte, D. Neshev, A. Tünnermann, and Y. S. Kivshar, "Diffraction control in periodically curved two-dimensional waveguide arrays," Opt. Express **15**, 9737 (2007).
14. S. Longhi and K. Staliunas, "Self-collimation and self-imaging effects in modulated waveguide arrays," Opt. Commun. **281**, 4343 (2008).
15. A. Szameit, I. L. Garanovich, M. Heinrich, A. A. Sukhorukov, F. Dreisow, T. Pertsch, S. Nolte, A. Tünnermann, and Y. S. Kivshar, "Polychromatic dynamic localization in curved photonic lattice," Nature Phys. **5**, 271 (2009).
16. K. Staliunas and C. Masoller, "Subdiffractive light in bi-periodic arrays of modulated fibers," Opt. Express **14**, 10669 (2006).
17. K. Staliunas and R. Herrero, "Nondiffractive propagation of light in photonic crystals," Phys. Rev. E **73**, 016601 (2006).
18. K. Staliunas, R. Herrero, G. J. de Valcarcel, "Arresting soliton collapse in two-dimensional nonlinear Schrödinger systems via spatiotemporal modulation of the external potential," Phys. Rev. A **75**, 011604(R) (2007).
19. A. Szameit, Y. V. Kartashov, F. Dreisow, M. Heinrich, T. Pertsch, S. Nolte, A. Tünnermann, V. A. Vysloukh, F. Lederer, and L. Torner, "Inhibition of light tunneling in waveguide arrays," Phys. Rev. Lett. **102**, 153901 (2009).
20. O. Peleg, G. Bartal, B. Freedman, O. Manela, M. Segev, and D. N. Christodoulides, "Conical Diffraction and Gap Solitons in Honeycomb Photonic Lattices," Phys. Rev. Lett. **98**, 103901 (2007).

# References without titles

1. F. Lederer, G. I. Stegeman, D. N. Christodoulides, G. Assanto, M. Segev, and Y. Silberberg, Phys. Rep. **463**, 1 (2008).
2. Y. V. Kartashov, V. A. Vysloukh, and L. Torner, Prog. Opt. **52**, 63 (2009).
3. M. J. Ablowitz and Z. H. Musslimani, Phys. Rev. Lett. **87**, 254102 (2001).
4. A. Szameit, I. L. Garanovich, M. Heinrich, A. Minovich, F. Dreisow, A. A. Sukhorukov, T. Pertsch, D. N. Neshev, S. Nolte, W. Krolikowski, A. Tünnermann, A. Mitchell, and Y. S. Kivshar, Phys. Rev. A **78**, 031801 (2008).
5. Y. V. Kartashov, L. Torner, and D. N. Christodoulides, Opt. Lett. **30**, 1378 (2005).
6. C. R. Rosberg, I. L. Garanovich, A. A. Sukhorukov, D. N. Neshev, W. Krolikowski, and Y. S. Kivshar, Opt. Lett. **31**, 1498 (2006).
7. Y. V. Kartashov, V. A. Vysloukh, and L. Torner, Phys. Rev. Lett. **99**, 233903 (2007).
8. K. Shandarova, C. E. Rüter, D. Kip, K. G. Makris, D. N. Christodoulides, O. Peleg, and M. Segev, Phys. Rev. Lett. **102**, 123905 (2009).
9. Y. V. Kartashov, L. Torner, and V. A. Vysloukh, Opt. Lett. **29**, 1102 (2004).
10. S. Longhi, M. Marangoni, M. Lobino, R. Ramponi, P. Laporta, E. Cianci, and V. Foglietti, Phys. Rev. Lett. **96**, 243901 (2006).
11. R. Iyer, J. S. Aitchison, J. Wan, M. M. Dignam, C. M. de Sterke, Opt. Express **15**, 3212 (2007).
12. F. Dreisow, M. Heinrich, A. Szameit, S. Doering, S. Nolte, A. Tünnermann, S. Fahr, and F. Lederer, Opt. Express **16**, 3474 (2008).
13. I. L. Garanovich, A. Szameit, A. A. Sukhorukov, T. Pertsch, W. Krolikowski, S. Nolte, D. Neshev, A. Tünnermann, and Y. S. Kivshar, Opt. Express **15**, 9737 (2007).
14. S. Longhi and K. Staliunas, Opt. Commun. **281**, 4343 (2008).
15. A. Szameit, I. L. Garanovich, M. Heinrich, A. A. Sukhorukov, F. Dreisow, T. Pertsch, S. Nolte, A. Tünnermann, and Y. S. Kivshar, Nature Phys. **5**, 271 (2009).
16. K. Staliunas and C. Masoller, Opt. Express **14**, 10669 (2006).
17. K. Staliunas and R. Herrero, Phys. Rev. E **73**, 016601 (2006).

18. K. Staliunas, R. Herrero, G. J. de Valcarcel, Phys. Rev. A **75**, 011604(R) (2007).
19. A. Szameit, Y. V. Kartashov, F. Dreisow, M. Heinrich, T. Pertsch, S. Nolte, A. Tünnermann, V. A. Vysloukh, F. Lederer, and L. Torner, Phys. Rev. Lett. **102**, 153901 (2009).
20. O. Peleg, G. Bartal, B. Freedman, O. Manela, M. Segev, and D. N. Christodoulides, Phys. Rev. Lett. **98**, 103901 (2007).

# Figure captions

Figure 1. (a) The refractive index distribution in a honeycomb array. (b) $U_\mathrm{m}$ versus modulation frequency at $\mu = 0.06$ and $A = 0.01$. (c) Resonance frequency versus depth $\mu$ of longitudinal modulation for $A = 0.01$. In all cases the resonance curves were calculated at a distance $\xi = 4T_\mathrm{b}$.

Figure 2. (a) $U_\mathrm{m}$ versus modulation frequency at $\mu = 0.10$ for $A = 0.01$ (curve 1) and $A = 0.25$ (curve 2). (b) $A^2$ versus width of the resonance curve defined at the level $U_\mathrm{m} = 0.7$ at $\mu = 0.10$. In all cases the resonance curves were calculated at a distance $\xi = 4T_\mathrm{b}$.

Figure 3. Field modulus distributions at (a),(d) $\xi = 0.25T_\mathrm{b}$, (b),(e) $0.50T_\mathrm{b}$, and (c),(f) $0.75T_\mathrm{b}$ for single-site excitation. Panels (a)-(c) correspond to an unmodulated array, while (d)-(f) correspond to a modulated array with $\mu = 0.1$, $\Omega = 6.4\Omega_\mathrm{b}$. In both cases the input amplitude is $A = 0.01$.

Figure 4. Field modulus distributions at (a),(d) $\xi = 0.25T_\mathrm{b}$, (b),(e) $0.50T_\mathrm{b}$, and (c),(f) $0.75T_\mathrm{b}$ for single-site excitation. In (a)-(c) the waveguides are modulated in-phase along the diagonal of the array, while in (d)-(f) they are modulated in-phase along the vertical axis. In all cases $\mu = 0.1$, $\Omega = 6.4\Omega_\mathrm{b}$, and the input amplitude is $A = 0.01$.

Figure 5. Field modulus distributions at (a),(d),(g) $\xi = 0.25T_\mathrm{b}$, (b),(e),(h) $0.50T_\mathrm{b}$, and (c),(f),(i) $0.75T_\mathrm{b}$ for vortex excitation with topological charge $m = 1$. Panels (a)-(c) correspond to unmodulated array, (d)-(i) correspond to a modulated array with $\mu = 0.1$, $\Omega = 6.4\Omega_\mathrm{b}$, while $A = 0.01$ in (a)-(f), and $A = 0.60$ in (g)-(i).

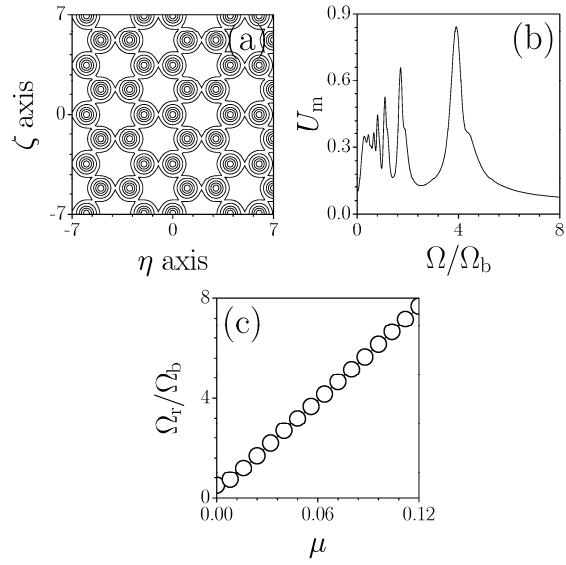

Figure 1.  (a) The refractive index distribution in a honeycomb array. (b) $U_\mathrm{m}$ versus modulation frequency at $\mu = 0.06$ and $A = 0.01$. (c) Resonance frequency versus depth $\mu$ of longitudinal modulation for $A = 0.01$. In all cases the resonance curves were calculated at a distance $\xi = 4T_\mathrm{b}$.

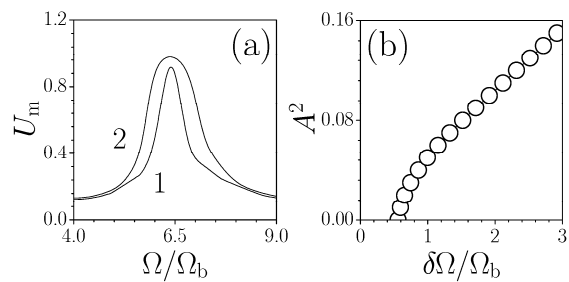

Figure 2. (a) $U_{\mathrm{m}}$ versus modulation frequency at $\mu = 0.10$ for $A = 0.01$ (curve 1) and $A = 0.25$ (curve 2). (b) $A^2$ versus width of the resonance curve defined at the level $U_{\mathrm{m}} = 0.7$ at $\mu = 0.10$. In all cases the resonance curves were calculated at a distance $\xi = 4T_{\mathrm{b}}$.

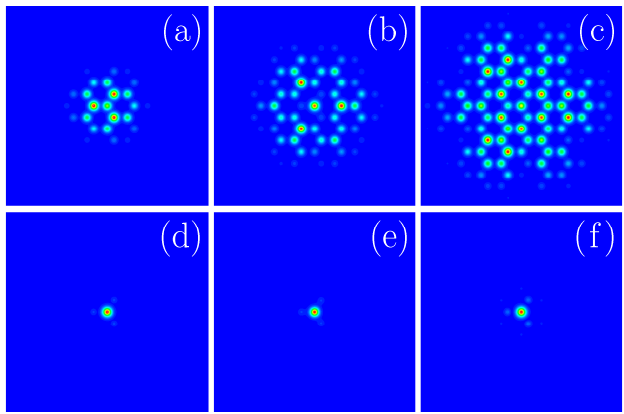

Figure 3. Field modulus distributions at (a),(d) $\xi = 0.25T_b$, (b),(e) $0.50T_b$, and (c),(f) $0.75T_b$ for single-site excitation. Panels (a)-(c) correspond to an unmodulated array, while (d)-(f) correspond to a modulated array with $\mu = 0.1$, $\Omega = 6.4\Omega_b$. In both cases the input amplitude is $A = 0.01$.

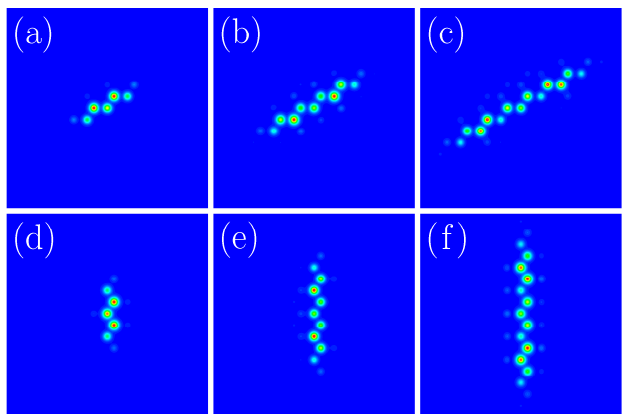

Figure 4. Field modulus distributions at (a),(d) $\xi = 0.25T_b$, (b),(e) $0.50T_b$, and (c),(f) $0.75T_b$ for single-site excitation. In (a)-(c) the waveguides are modulated in-phase along the diagonal of the array, while in (d)-(f) they are modulated in-phase along the vertical axis. In all cases $\mu = 0.1$, $\Omega = 6.4\Omega_b$, and the input amplitude is $A = 0.01$.

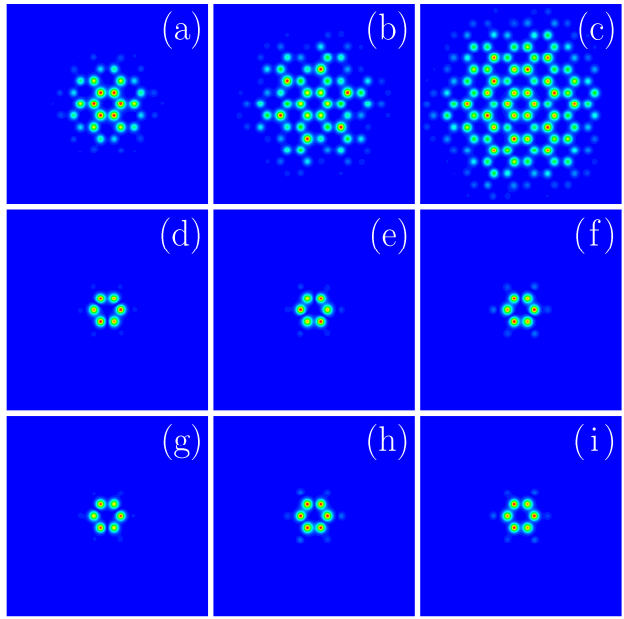

Figure 5. Field modulus distributions at (a),(d),(g) $\xi = 0.25T_{\rm b}$, (b),(e),(h) $0.50T_{\rm b}$, and (c),(f),(i) $0.75T_{\rm b}$ for vortex excitation with topological charge $m = 1$. Panels (a)-(c) correspond to unmodulated array, (d)-(i) correspond to a modulated array with $\mu = 0.1$, $\Omega = 6.4\Omega_{\rm b}$, while $A = 0.01$ in (a)-(f), and $A = 0.60$ in (g)-(i).